\documentclass[12pt]{iopart}
\usepackage{iopams}
\usepackage{epsfig}

\begin{document}

\def\be{\begin{equation}}
\def\ee{\end{equation}}
\def\bea{\begin{eqnarray}}
\def\eea{\end{eqnarray}}
\def\mt{\widetilde{m}_1}
\def\ve{\varepsilon}
\def\d{\delta}
\def\k{\kappa}
\def\b{\beta}
\def\NO{\nonumber}
\def\mb{\overline{m}}

\def\pl#1#2#3{Phys.~Lett.~{\bf B {#1}} ({#2}) #3}
\def\np#1#2#3{Nucl.~Phys.~{\bf B {#1}} ({#2}) #3}
\def\prl#1#2#3{Phys.~Rev.~Lett.~{\bf #1} ({#2}) #3}
\def\pr#1#2#3{Phys.~Rev.~{\bf D {#1}} ({#2}) #3}
\def\zp#1#2#3{Z.~Phys.~{\bf C {#1}} ({#2}) #3}
\def\cqg#1#2#3{Class.~and Quantum Grav.~{\bf {#1}} ({#2}) #3}
\def\cmp#1#2#3{Commun.~Math.~Phys.~{\bf {#1}} ({#2}) #3}
\def\jmp#1#2#3{J.~Math.~Phys.~{\bf {#1}} ({#2}) #3}
\def\ap#1#2#3{Ann.~of Phys.~{\bf {#1}} ({#2}) #3}
\def\prep#1#2#3{Phys.~Rep.~{\bf {#1}C} ({#2}) #3}
\def\ptp#1#2#3{Progr.~Theor.~Phys.~{\bf {#1}} ({#2}) #3}
\def\ijmp#1#2#3{Int.~J.~Mod.~Phys.~{\bf A {#1}} ({#2}) #3}
\def\mpl#1#2#3{Mod.~Phys.~Lett.~{\bf A {#1}} ({#2}) #3}
\def\nc#1#2#3{Nuovo Cim.~{\bf {#1}} ({#2}) #3}
\def\ibid#1#2#3{{\it ibid.}~{\bf {#1}} ({#2}) #3}

\title[Leptogenesis]{Some Aspects of Thermal Leptogenesis}

\author{W.~Buchm\"uller$^{\rm a}$, P.~Di Bari$^{\rm b}$ and M.~Pl\"umacher$^{\rm c}$}

\address{$^{\rm a}$ Deutsches Elektronen-Synchrotron DESY, 22603 Hamburg, Germany}

\address{$^{\rm b}$ IFAE, Universitat Aut\`onoma de Barcelona,
08193 Bellaterra (Barcelona), Spain}

\address{$^{\rm c}$ Department of Physics, CERN, Theory Division,
1211 Geneva 23, Switzerland}

\begin{abstract}
Properties of neutrinos may be the origin of the matter-antimatter
asymmetry of the universe. In the seesaw model for neutrino masses
this leads to important constraints on the properties of light and
heavy neutrinos. In particular, an upper bound on the light neutrino
masses of $0.1\,$eV can be derived.  We review the present status of
thermal leptogenesis with emphasis on the theoretical uncertainties
and discuss some implications for lepton and quark mass hierarchies,
$C\!P$ violation and dark matter. We also comment on the `leptogenesis
conspiracy', the remarkable fact that neutrino masses may lie in the
range where leptogenesis works best.
\end{abstract}



\maketitle

\section{Introduction}

One of the great challenges of modern particle physics and cosmology
is to explain the excess of matter over anti-matter observed in the
universe. This baryon asymmetry is conveniently expressed as the
ratio of baryon minus anti-baryon density to the photon density
and has recently been measured to a high degree of accuracy by
observations of the cosmic microwave background (CMB) \cite{WMAP}
combined with measurements of large scale structures of the universe
\cite{SDSS}:
\begin{equation}\label{etaBobs}
  \eta_B^{\rm CMB}=(6.3\pm0.3)\times 10^{-10}\;.
  \label{etaBCMB}
\end{equation}
In a complete cosmological model this baryon asymmetry has to be
dynamically generated during the evolution of the universe in
the hot and dense phase shortly after the big bang. This is possible
if the particle interactions violate baryon number ($B$), charge
conjugation ($C$) and the combined charge and parity conjugation ($C\!P$),
and if the expansion of the universe leads to a deviation from
thermal equilibrium \cite{Sakharov}.

All these ingredients are present in the Standard Model (SM) of
particle interactions. However, the baryon asymmetry that can be
generated in the SM falls far short of observations, i.e.\ an
extended model of particle interactions has to be considered.
The observation of neutrino masses also requires an extension
of the SM and lepton number ($L$) violating interactions that
are introduced in the seesaw model of neutrinos masses \cite{yan79,grs79}
can naturally give rise to the observed baryon asymmetry in the
leptogenesis scenario \cite{FY86} that is the topic of this article.

The rather suprising fact that lepton number violating interactions
can give rise to a baryon asymmetry is due to a deep connection
between baryon and lepton number in the SM, as discussed in section
\ref{blabla}. In section \ref{basics} we present the basic mechanism
of leptogenesis and introduce some notations that are used in the
following. The quantitative solution of the Boltzmann equations and
the corresponding bounds on light and heavy neutrino masses are 
described in section 4. Here we also comment on the {\it leptogenesis
conspiracy}, the remarkable fact that neutrino masses may lie
in the range where leptogenesis works best. In section 5 the connection
between low and high energy $C\!P$ violation and implications of leptogenesis
for the heavy neutrino mass spectrum are briefly discussed. Section 6 
deals with the implications of leptogenesis for dark matter.

\section{Baryon and lepton number violation in the SM \label{blabla}}

In the SM both baryon and lepton number are classically conserved,
since they are protected by global Abelian symmetries. However,
due to the chiral nature of weak interactions, these symmetries are
anomalous and are violated at the quantum level \cite{tHooft}.
This is related to the non trivial vacuum structure of non-Abelian
gauge theories, like the SM. Neglecting fermion masses, there are
an infinite number of degenerate ground states whose vacuum field
configurations have different topological charges, or Chern-Simons numbers 
$N_{CS}$ \cite{topo,km84}. In the electroweak sector of the SM a change in
the topological charge, i.e.\ a transition from one vacuum to
another one, corresponds to a change in baryon and lepton numbers,
\begin{equation}
  \Delta B=\Delta L= n_f \Delta N_{CS}\;,
  \label{DeltaNCS}
\end{equation}
where $n_f$ is the number of generations of quarks and leptons,
i.e.\ $n_f=3$ in the SM. Note that, although both baryon and lepton
number are violated, the linear combination $B-L$ is still conserved
at the quantum level.

At low temperatures, when the electroweak symmetry is broken,
the different vacua are separated by a potential barrier, whose
height is determined by the vacuum expectation value (VEV) of the
Higgs field, $v=\langle\phi\rangle$, i.e.\ the scale of electroweak
symmetry breaking.
Hence, processes changing the topological charge are tunneling
processes whose rate is unobservably small, due to the smallness
of the electroweak coupling constant. In the low temperature
regime being probed at accelerator experiments,
$B$ and $L$ are therefore conserved to a very good approximation,
in accord with experimental observations (cf.~\cite{Rubakov}).

\begin{figure}[t]
\centerline{\psfig{file=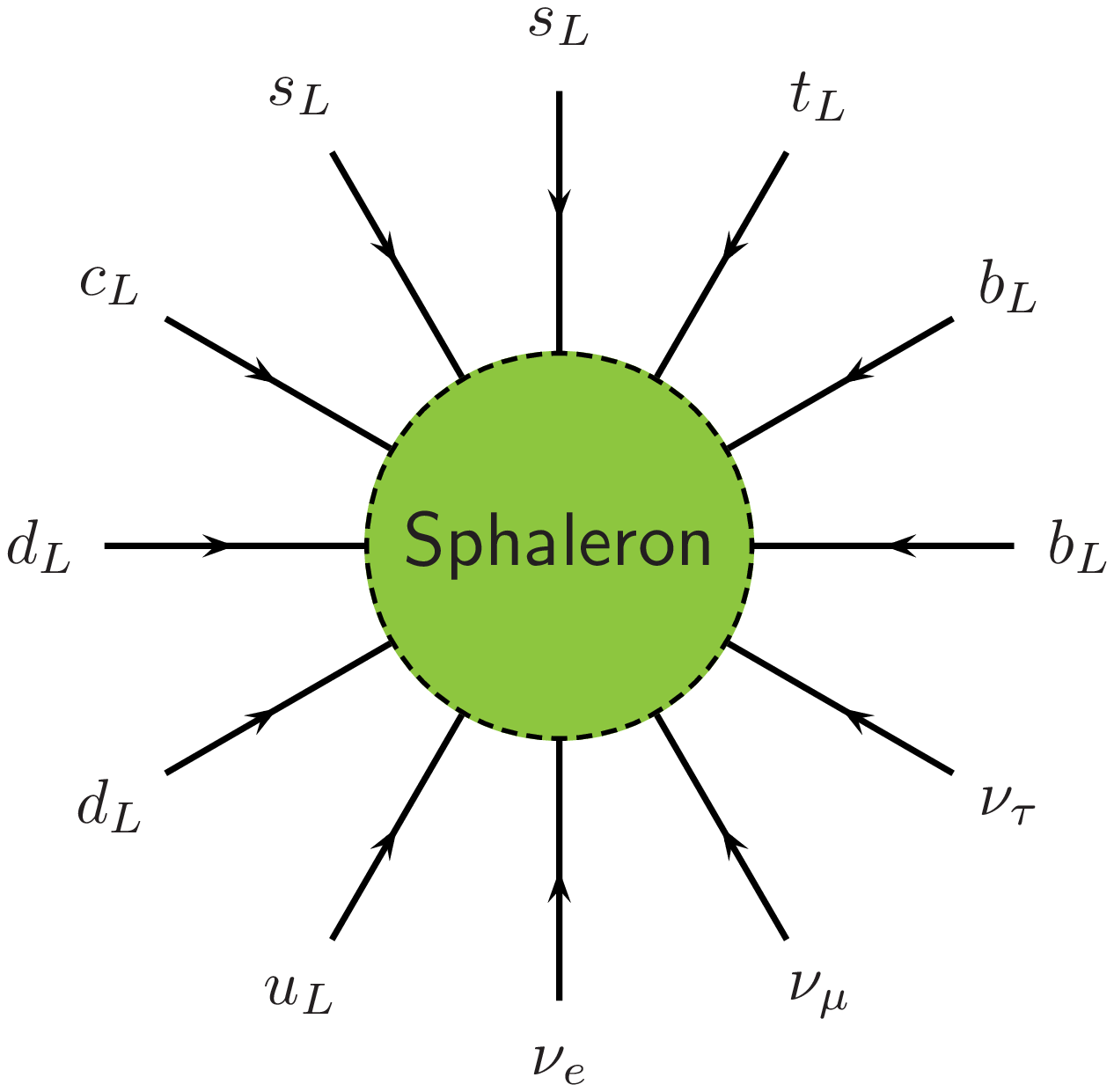,height=7cm,width=7cm}}
\caption{\it One of the 12-fermion processes which are in thermal
equilibrium in the high-temperature phase of the standard model.
\label{fig_sphal}}
\end{figure}

When the standard model particles form a heat bath of temperature $T$
the situation changes. At high temperatures, $T\geq T_{EW}\sim 100\;$GeV,
the Higgs VEV `evaporates', leading to a restoration of the electroweak
symmetry and the disappearance of the potential barriers separating
the different vacua. $B$ and $L$ violating transitions are then no
longer suppressed \cite{Sphaleron}.

The rate at which these processes occur is related to the free energy
of field configurations which carry topological charge. In the
electroweak part of the SM these so-called sphaleron processes
lead to an effective interaction of all left-handed fermions \cite{tHooft}
(cf.\ Fig.~\ref{fig_sphal}),
\begin{equation}
  O_{B+L} = \prod\limits_{i}\left(q_{Li}q_{Li}q_{Li}l_{Li}\right)\,,
\end{equation}
which indeed violates both baryon and lepton number by three units
but conserves the combination $B-L$, in accord with Eq.~(\ref{DeltaNCS}).

The sphaleron transition rate in the symmetric phase of the SM has
been evaluated by combining an analytical resummation with numerical
lattice techniques \cite{Dietrich}. The result is that sphaleron
processes are in thermal equilibrium for temperatures in the range
\begin{equation}
  100\,{\rm GeV}\lesssim T \lesssim 10^{12}\,{\rm GeV}\;.
\end{equation}
These processes have a profound effect on the generation of the
cosmological baryon asymmetry.  Eq.~(\ref{DeltaNCS}) suggests that any
$B+L$ asymmetry generated at temperatures $T>T_{EW}$, will be washed out.
However, since only left-handed fields couple to sphalerons, a non-zero
value of $B+L$ can persist in the high-temperature, symmetric phase if
there exists a non-vanishing $B-L$ asymmetry. An analysis of the chemical
potentials of all particle species in the high-temperature phase yields
the following relation between the baryon asymmetry $\eta_B$ and the
corresponding $L$ and $B-L$ asymmetries $\eta_L$ and $\eta_{B-L}$,
respectively \cite{ht90},
\begin{equation}
  \eta_B\ =\ a_{sph}\ \eta_{B-L}\ =\ {a_{sph}\over a_{sph}-1}\ \eta_L\;.
  \label{basic}
\end{equation}
Here $a_{sph}$ is a number ${\cal O}(1)$. In the SM with three
generations and one Higgs doublet one has $a_{sph}=28/79$.

\section{Leptogenesis \label{basics}}

\subsection{A qualitative overview \label{quali}}

The deep connection between baryon and lepton number in the early
universe has led to the realization that lepton number violating
processes, whose presence is predicted by the seesaw model
for light neutrino masses \cite{yan79,grs79}, can be responsible for the
observed cosmological baryon asymmetry.

In the seesaw model the smallness of light neutrino masses
is explained through the mixing of left-handed neutrinos with
right-handed neutrinos $\nu_R$ which are not present in the
SM but are predicted in certain models of grand unification. The
interactions of the SM are supplemented by the following Yukawa
couplings of neutrinos,
\begin{equation}
  {\cal L}_Y= \overline{l_L}\,h\,\nu_R\,\phi +
             \overline{\nu_R^c}\,M\,\nu_R + {\rm h.c.}\;,
  \label{lagrangian}
\end{equation}
where $M$ is the Majorana mass matrix of the right-handed neutrinos,
and the Yukawa couplings $h$ yield the Dirac neutrino mass matrix
$m_D = hv$ after spontaneous breaking of the electroweak symmetry.
Since the Majorana mass matrix $M$ is independent of electroweak symmetry
breaking, one can have $M \gg m_D$, which leads to the mass eigenstates
\begin{equation}
\nu \simeq \nu_L + \nu_L^c = \nu^c\;, \quad N \simeq\nu_R + \nu_R^c = N^c\;,
\end{equation}
with heavy neutrino masses $M$ and the light neutrino masses
\begin{equation}\label{wettss}
m_\nu = - m_D {1\over M} m_D^T\;.
\end{equation}

The heavy neutrinos are unstable and decay through their Yukawa
couplings into SM lepton and Higgs doublets. Due to their Majorana
nature, the heavy neutrinos $N$ can decay both into leptons and
anti-leptons, i.e.\ lepton number is violated in these decays. In
conjunction with $B+L$ violating sphaleron transitions this leads to
the required non-conservation of baryon number. Further, violation of
$C$ and $C\!P$ comes about since the Yukawa couplings $h$ are, in general,
complex, thereby making possible the generation of a non-vanishing
baryon asymmetry in these decays.

A further complication is that the heavy neutrinos also mediate
lepton number violating scattering processes which can erase any
lepton asymmetry \cite{fy90,ht90}. However, the interaction rates for these 
processes
are suppressed by the large mass of the heavy neutrinos if the
temperature $T$ is smaller than their mass.  Hence, the lepton
asymmetries produced in decays of the heavier neutrinos
$N_{2,3}$ will be erased by lepton number violating scatterings
mediated by the lightest of the heavy neutrinos, $N_1$. Therefore, in
the simplest case of hierarchical heavy neutrino masses, $M_1\ll
M_{2,3}$, only decays of $N_1$ can potentially explain the observed
baryon asymmetry.

The required deviation from thermal equilibrium is provided by the
expansion of the universe. When the universe has cooled down to a
temperature of order the heavy neutrino mass $M_1$, the
equilibrium number density becomes exponentially suppressed.  If the
neutrinos are sufficiently weakly coupled they are not able to follow
the rapid change of the equilibrium particle distribution once the
temperature falls below their mass. Hence, the deviation from thermal
equilibrium consists in a too large number density of heavy neutrinos
compared to the equilibrium density \cite{kt}.  Technically this
requires the total decay width of $N_1$ to be smaller than the
expansion rate, the Hubble parameter $H$, at the time of decay, i.e.\
when $T\sim M_1$. This is the case if the {\em effective neutrino
mass}, defined as
\begin{equation}
  \widetilde{m}_1={(m_D^{\dagger} m_D)_{11}\over M_1}\;,
\end{equation}
is smaller than the {\em equilibrium neutrino mass}
\begin{equation}
m_* = {16\pi^{5/2}\over 3\sqrt{5}} g_*^{1/2} {v^2\over M_{\rm p}} 
\simeq 10^{-3}~\mbox{eV}\;,
\end{equation}
where we have used $M_{\rm p} = 1.2\times 10^{19}$~GeV and $g_* = 106.75$ as 
effective number of relativistic degrees of freedom in the plasma. 
The effective neutrino mass $\widetilde{m}_1$ is a measure of
the strength of the coupling of $N_1$ to the thermal bath.

In order to see whether this mechanism of leptogenesis \cite{FY86}
can indeed explain the observed baryon asymmetry a careful
numerical study is needed. As we shall see, successful leptogenesis is
possible for $\mt < m_*$ as well as $\mt > m_*$.
A quantitative description of this
non-equilibrium process is obtained by means of kinetic equations.

\subsection{Boltzmann equations \label{Boltzmann}}
The evolution of particle number densities in the early universe
is influenced not only by interactions but also by the expansion
of the universe. It is convenient to scale out the expansion by
considering the particle number $N_X$ in some comoving volume
element instead of the number density $n_X$. For definiteness,
we choose the comoving volume $R_*(t)^3$ which contains one
photon at a time $t_*$ before the onset of leptogenesis,
\begin{equation}
  N_X(t)=n_X(t)\,R_*(t)^3\;.
\end{equation}
The final baryon asymmetry is expressed in terms of the
baryon-to-photon ratio $\eta_B$, to be compared with the observed
value $\eta_B^{\rm CMB}$ (cf.~Eq.~(\ref{etaBCMB})). This is related
to the $B-L$ asymmetry in a comoving volume element by
\begin{equation}\label{etaB}
  \eta_B={a_{sph}\over f} N_{B-L}^{\rm f},
\end{equation}
where $f=2387/86$ is the dilution factor due to the production
of photons from the onset of leptogenesis until recombination,
assuming the standard isentropic expansion of the universe.

Further, it is convenient to replace time $t$ by $z=M_1/T$,
where $M_1$ is the mass of the decaying neutrino. This is
possible, since in a radiation dominated universe both variables
are related by the expansion rate of the universe, the Hubble
parameter $H$,
\begin{equation}
  H=2t=\sqrt{4\pi^3g_*\over45}\,{M_1^2\over M_{\rm p}}\,{1\over z^2}\;.
\end{equation}

In the simplest case of a hierarchical mass spectrum of right-handed
neutrinos, $M_1\ll M_{2,3}$, a numerical description of leptogenesis
is provided by a set of two coupled differential equations
\cite{Luty,bdp02},
\begin{eqnarray}
 {dN_{N_1}\over dz} &=& -(D+S)\,\left(N_{N_1}-N_{N_1}^{\rm eq}\right)
                        \label{B1}\;,\\[1ex]
 {dN_{B-L}\over dz} &=& -\varepsilon_1\,D\,\left(N_{N_1}-N_{N_1}^{\rm eq}\right)
                        -W\,N_{B-L}
                        \label{B2}\;,
\end{eqnarray}
where the terms on the right-hand side describe the effects of particle
interactions. There are four classes of processes which contribute:
decays, inverse decays, $\Delta L=1$ scatterings and $\Delta L=2$
processes mediated by heavy neutrinos. The term $D$ accounts for
decays and inverse decays, while the scattering term $S$ represents
the $\Delta L=1$ scatterings. Decays also yield the source term
for the generation of the $B-L$ asymmetry, the first term in Eq.~(\ref{B2}),
while all other processes contribute to the total washout term $W$ which
competes with the decay source term. Note that, in thermal equilibrium,
$N_{N_1}=N_{N_1}^{\rm eq}$, no $B-L$ asymmetry can be generated,
illustrating the need for a deviation from thermal equilibrium.
The amount of $B-L$ asymmetry being produced by the source term
is controlled by the $C\!P$ asymmetry $\varepsilon_1$ in the decay
of $N_1$.

In order to understand the dependence of the solutions on the
neutrino parameters, it is crucial to note that the interactions
terms $D$ and $S$ as well as the contribution from $N_1$ exchange
to $W$ are all proportional to the effective neutrino mass,
\begin{equation}
  D,\; S,\; W_1\;\propto {M_{\rm p}\widetilde{m}_1\over v^2}\;,
\end{equation}
whereas the strength of the remaining contribution to the washout term,
$\Delta W$, is determined by $\overline{m}^2=
m_1^2+m_2^2+m_3^2$, the sum over the light neutrino
masses squared,
\begin{equation}\label{DW}
  \Delta W \propto {M_{\rm p}M_1\overline{m}^2\over v^4}\;.
\end{equation}

If one assumes a vanishing initial $B-L$ asymmetry before the onset
of leptogenesis, i.e.\ at $z\ll1$, the solution for $N_{B-L}$ has
the simple form
\begin{equation}\label{NBmL}
  N_{B-L}(z)=-{3\over4}\,\varepsilon_1\,
             \kappa(z;\widetilde{m}_1,M_1\overline{m}^2)\;,
\end{equation}
where we have introduced the {\em efficiency factor} $\kappa$ \cite{strumia}
which does not depend on the $C\!P$ asymmetry $\varepsilon_1$ and parametrizes
the effect of scattering and decay processes. It is given by the
following integral expression:
\begin{equation}\label{kz}
  \kappa(z)={4\over3}\int\limits_{z_i}^zdz'\,D\,
            \left(N_{N_1}-N_{N_1}^{\rm eq}\right)
            e^{-\int_{z'}^z dz''\,W(z'')}\;.
\end{equation}
It is normalized in such a way that its final value
$\kappa_f=\kappa(\infty)$ approaches one in the limit of thermal
initial abundance of heavy neutrinos, $N_{N_1}(z\ll1)=N_{N_1}^{\rm eq}=3/4$
and no washout, $W=0$.

\begin{figure}[t]
\centerline{\psfig{file=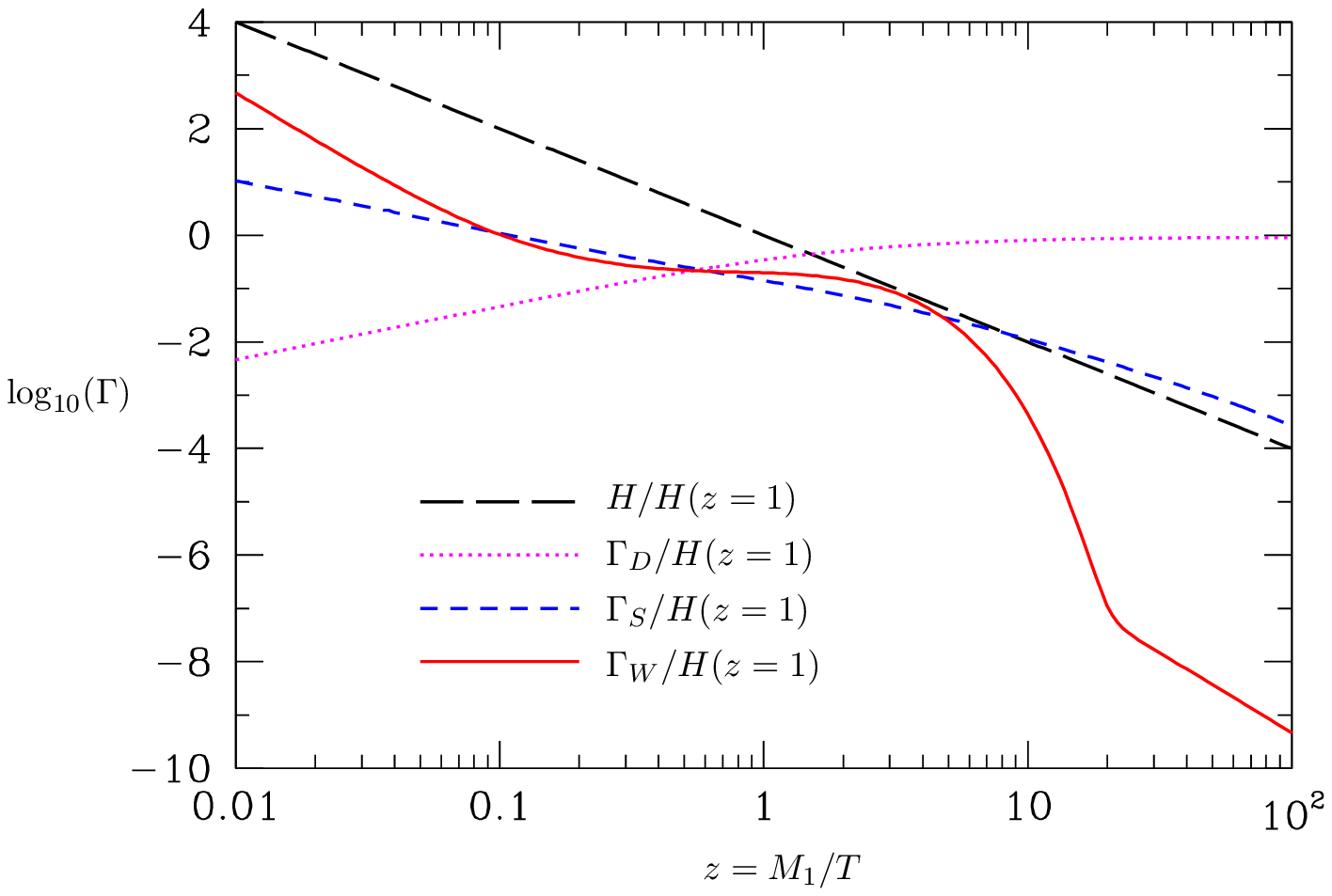,width=14cm}}
\caption{The  expansion rate of the universe and the three interaction
rates nomalized to the expansion rate at $z=1$ for a typical choice
of parameters, $M_1=10^{10}\,$GeV, $\widetilde{m}_1=10^{-3}\,$eV
and $\overline{m}=0.05\,$eV.}
\label{Rates}
\end{figure}
\begin{figure}[ht]
\centerline{\psfig{file=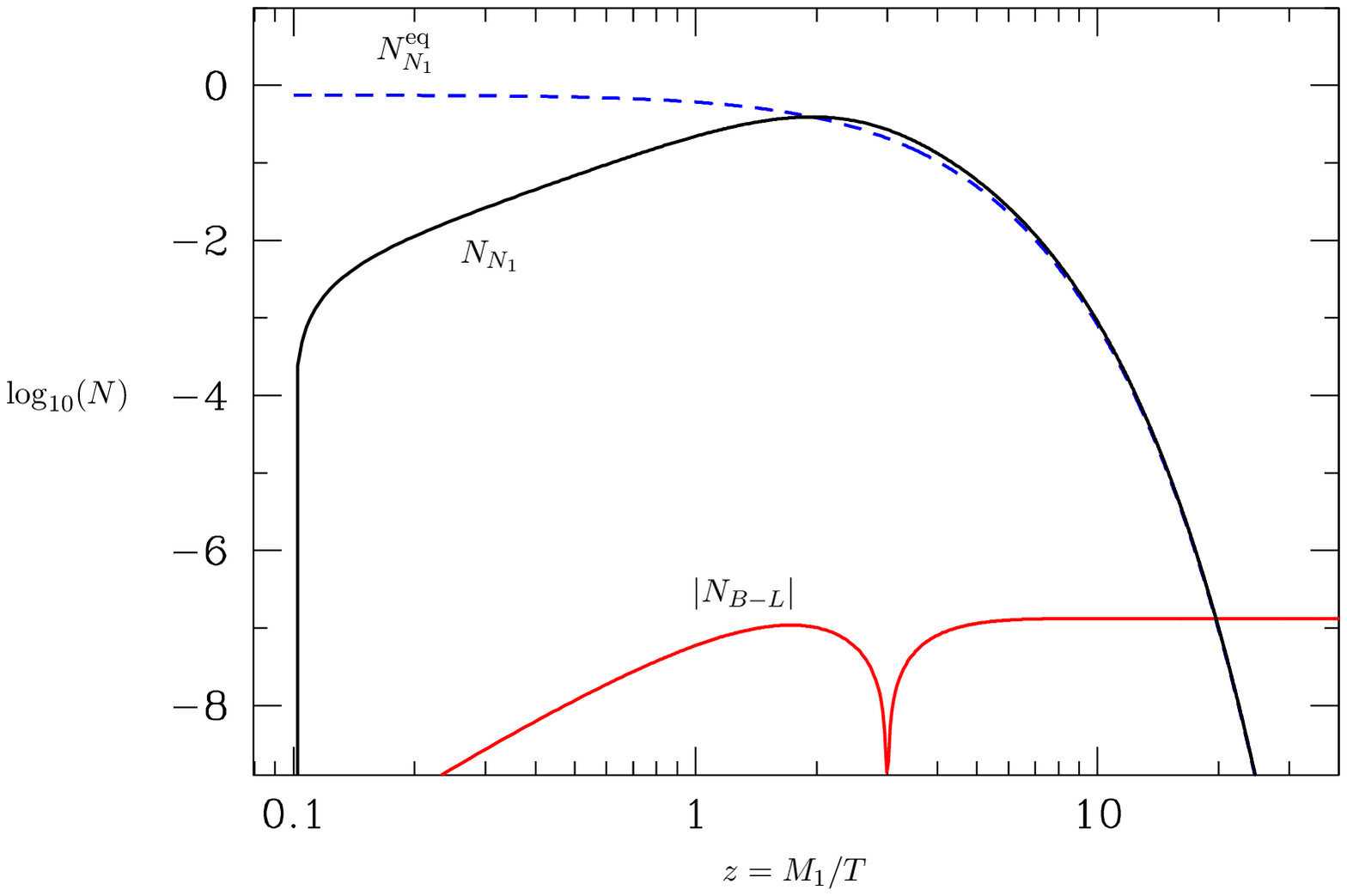,width=14cm}}
\caption{The evolution of the $N_1$ abundance and the $B-L$ asymmetry
for the same choice of parameters as in Fig.~\ref{Rates} and
$\varepsilon_1=10^{-6}$.}
\vspace{-5mm}
\label{B}
\end{figure}

As an example, the three interaction rates,
$(\Gamma_D,\Gamma_S,\Gamma_W)=H\,z\,(D,S,W)$, and the Hubble parameter
are shown in Fig.~\ref{Rates} for a typical choice of parameters,
$M_1=10^{10}\,$GeV, $\widetilde{m}_1=10^{-3}\,$eV and
$\overline{m}=0.05\,$eV. As can be seen from the figure, the
out-of-equilibrium condition $\widetilde{m}_1\lesssim 10^{-3}\,$eV
is fulfilled at $z=1$, i.e.\ all interaction rates are smaller
than the Hubble parameter.

The corresponding evolution of the $N_1$ abundance and the $B-L$ asymmetry
is shown in Fig.~\ref{B}, starting at $z=0.1$ with a vanishing initial
$N_1$ abundance. Although the neutrino production rates $\Gamma_D$ and
$\Gamma_S$ are not in thermal equilibrium, they are still strong enough
to produce a non-vanishing abundance of neutrinos at $z\lesssim1$ and the
equilibrium distribution is reached at $z\sim2$. The required deviation
from thermal equilibrium can clearly be seen as a small over-abundance of
neutrinos for $z\gtrsim2$. However, the decay rate $\Gamma_D$ also comes
into thermal equilibrium at $z\sim2$ leading to a rapid decay of the
$N_1$ abundance and the production of a non-vanishing $B-L$ asymmetry.
The change in sign in the asymmetry at $z\sim3$ is due to the fact
that the source term in eq.~(\ref{B2}) changes sign once the $N_1$
abundance becomes larger than the equilibrium abundance, i.e.\
neutrino production processes at $z\lesssim2$ lead to a `wrong sign'
asymmetry that partially cancels against the asymmetry produced in
the $N_1$ decays at $z\gtrsim2$.

\section{CMB constraints on neutrino masses}

We are now ready to discuss the implications of thermal leptogenesis
for light and heavy neutrino masses.
Combining Eqs.~(\ref{etaB}) and (\ref{NBmL}) one obtains
\begin{equation}\label{etak}
   \eta_B= -d\,\,\ve_1\,
           \kappa_{\rm f}(\mt,M_1\overline{m}^2)\; ,
\end{equation}
where $d=3\,a_{\rm sph}/(4\, f) \simeq 10^{-2}$. Hence, determining 
the amount of baryon
asymmetry produced in leptogenesis requires the calculation of both the final
efficiency factor $\k_{\rm f}$ and the $C\!P$ asymmetry $\ve_1$. A
comparison with the observed value (cf.~Eq.~(\ref{etaBobs})) then
allows to place stringent constraints on the involved seesaw
parameters and, remarkably, on light and heavy neutrino masses.

\subsection{Final efficiency factor}

Starting from Eq.~(\ref{kz}) and assuming a high initial temperature,
$z_{i}=M_1/T_{\rm i}\ll 1$, the final efficiency factor can be
calculated analytically \cite{kt,bdp04}.  For values
$M_1\,\mb^2 \ll 10^{14}\,{\rm GeV}\,[\mb/(0.05\,{\rm eV})]^2$
the washout term $\Delta W$ can be neglected and the final efficiency
factor depends only on the effective neutrino mass $\mt$.

\begin{figure}[t]
\centerline{\psfig{file=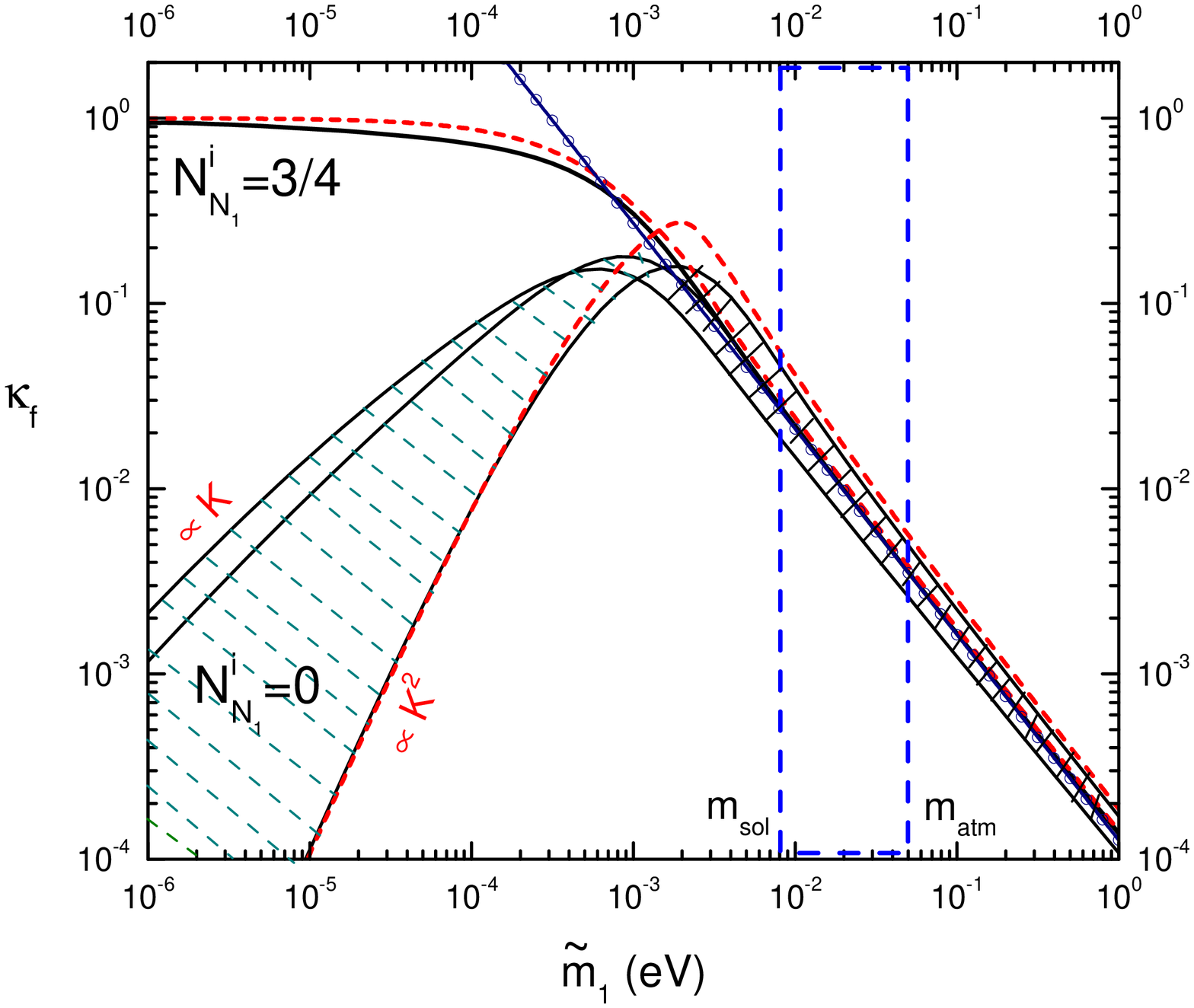,height=11cm,width=16cm}}
\vspace{-5mm}
\caption{\small Final efficiency factor when the washout term $\Delta W$
is neglected.}
\label{Skf}
\end{figure}

The results for the efficiency factor are summarized in
Fig.~\ref{Skf}. Two different regimes can clearly be distinguished.
In the {\em weak washout regime}, $\widetilde{m}_1\ll
10^{-3}\,{\rm eV}$, the results strongly depend on the initial
conditions and on theoretical assumptions, i.e.\ in that case
predictions are strongly model dependent and affected by large
theoretical uncertainties. On the other hand, in the {\em strong
washout regime}, $\widetilde{m}_1\gg 10^{-3}\,{\rm eV}$, the
dependence on the initial conditions is practically negligible and the
theoretical uncertainties are small such that the final asymmetry can
be predicted within $\sim 50\%$.

In both cases very precise analytical approximations for the final
efficiency factor can be obtained. It is instructive to start with a
simplified picture where the scattering term $S$ is neglected and only
decays and inverse decays contribute. The {\it decay parameter},
\be\label{Kdef}
K\equiv{\Gamma_D(z=\infty)\over H(z=1)}={\widetilde{m}_1\over m_*} \, ,
\ee
controls whether $N_1$ decays are in thermal equilibirium or not. Here
$\Gamma_D(z=\infty)$ is the $N_1$ decay width and $m_*$ marks
the boundary between the weak and strong washout regimes, as
discussed in Section \ref{quali}.

In the {\em strong washout regime} inverse decays rapidly thermalize the 
heavy neutrinos $N_1$ and the washout due to decays and inverse decays is
strong enough to destroy an initial asymmetry that may have been
present before the onset of leptogenesis \cite{bdp03}, leading to a
negligible dependence on initial conditions.  Further, the integrand
in Eq.~(\ref{kz}) is peaked at a value $z_B \gg 1$, implying that the
final asymmetry is produced around a well defined {\em temperature of
baryogenesis} $T_B=M_1/z_B\ll M_1$, where the heavy neutrinos are fully 
non-relativistic. This contributes to reducing the theoretical
uncertainties in the strong washout regime, since it has been shown
that the Boltzmann equations employed here can be derived from a fully
consistent quantum mechanical treatment in terms of Kadanoff-Baym
equations in the non-relativistic limit \cite{bf00}.

In the strong washout regime the $N_1$ abundance closely
tracks the equilibrium abundance and a simple expression for the
final efficiency factor can be obtained,
\begin{equation}\label{kfth}
\k_f(K) \simeq {2\over K\,z_B(K)}\,
\left(1-e^{-{1\over2}\,K\,z_B(K)}\right) \; .
\end{equation}
Assuming thermal initial $N_1$ abundance, i.e.
$N_{N_1}^{\rm i}=3/4$, this expression also reproduces the correct
asymptotical limit in the weak washout regime, $K\ll 1$. In Fig.~\ref{Skf}
this analytical result is represented by the short-dashed line which
has to be compared to the numerical results, given by the solid lines.

In the weak washout regime the calculation is more involved if one
starts from a vanishing initial $N_1$ abundance. Indeed, in this
case the $N_1$ production by inverse decays leads to a
negative contribution to the efficiency factor, corresponding to a
`wrong-sign' asymmetry, as discussed in Section \ref{Boltzmann}. The
final efficiency factor is thus the sum of a negative contribution,
$\kappa^{-}_{\rm f}(K)$, and a positive one, $\kappa^{+}_{\rm f}(K)$.
A very good approximation for these contributions that interpolates
between the weak and strong washout regimes is given by
\begin{eqnarray}
  \kappa^{-}_{\rm f}(K) &=&
      -2\ e^{-{2\over 3}\,N(K)}
      \left(e^{{2\over 3} \overline{N}(K)} - 1 \right)
      \label{kf-} \;,\\
  \kappa^{+}_{\rm f}(K) &=&
      {2\over z_B(K)\,K}
      \left(1-e^{-{2\over 3} z_B(K)\,K \overline{N}(K)}\right) \; .
      \label{kf+}
\end{eqnarray}
Here $N(K)\simeq (9\pi/16)K$ is the maximal $N_1$ number density
being produced in the weak washout regime and $\overline{N}(K)$
interpolates between $N(K)$ and the maximal number density $N_{\rm
eq}=3/4$ in the strong washout regime.  For large $K$ the negative
contribution is suppressed, while the positive one asymptotically
approaches Eq.~(\ref{kfth}). On the other hand, in the weak washout
regime the positive and negative contributions cancel each other to
leading order in $K$, i.e.\ the total efficiency factor is of order
$K^2$ \cite{fry}, as shown in Fig.~\ref{Skf}, where the short-dashed line 
again corresponds to the analytical solution and the solid one to the
numerical integration of the Boltzmann equations.

This cancellation of the leading order contributions to the final
efficiency factor in the weak washout regime no longer occurs when
the scattering term $S$ is taken into account. Indeed, these
scattering processes enhance the $N_1$ production thereby
giving rise to an additional positive contribution to the efficiency
factor. On the other hand, these scatterings are $C\!P$-conserving,
i.e. they do not contribute to the negative part of the efficiency
factor, as long as the contribution of $\Delta L=1$ scatterings to the
washout term is negligible which is always the case in the weak
washout regime. In this way scatterings can greatly enhance the final
efficiency factor in the weak washout regime \cite{michele}. The
drawback is that the result is very sensitive to different
approximations being used in the computation of the scattering rates
and different results, ranging from the case where scatterings are
negligible to a behaviour $\k_{\rm f}\propto K$, have been
obtained. Potentially important effects that have recently been
discussed but are presently controversial include scattering processes
involving gauge bosons \cite{pilaftsis,gnx04} and thermal
corrections to the decay and scattering rates \cite{covi,gnx04}.
The range of different results is represented in Fig.~\ref{Skf} by
the hatched region. An additional uncertainty in the weak washout
regime comes about due to the dependence of the final results on the
initial $N_1$ abundance and a possible initial asymmetry
created before the onset of leptogenesis.

The situation is very different in the strong washout regime. The
final efficiency factor is not sensitive to the neutrino production
since a thermal neutrino distribution is always reached at high
temperatures, $z\ll1$. However, $\Delta L=1$ scatterings also
contribute to the washout term but their effect is small and thus the
theoretical uncertainty arising from these scattering processes is
not larger than about $50\%$. This uncertainty for values $\mt > m_* \simeq
10^{-3}\,{\rm eV}$ is again indicated by the hatched region in
Fig.~\ref{Skf}.  A much more important source of uncertainties are
{\em spectator processes} \cite{bp01}, which can change the
produced asymmetry by a factor of order one.

In the strong washout regime the highest efficiency is reached when
scatterings are neglected and only decays and inverse decays are taken
into consideration, which approximately corresponds to the results
obtained in \cite{gnx04}. In this parameter range the final efficiency
factor is given, within theoretical uncertainties, by the simple
power law \cite{bdp04}
\be\label{plaw}
  \k_{\rm f}=(2\pm1)\,10^{-2}\,
             \left({0.01\,{\rm eV}\over\mt}\right)^{1.1\pm 0.1}\;.
\ee
The only model independent information we have on $\widetilde{m}_1$ is
that it has to be larger than the smallest neutrino mass
$m_1$ \cite{fhy02}. However, a situation where $m_1<\mt\ll m_{\rm sol}$
is rather artificial within neutrino mass models and the leptogenesis
predictions are then very model dependent in the weak washout regime. In
typical neutrino mass models values of $\widetilde{m}_1$ are usually
in the mass range suggested by neutrino oscillations. It is remarkable
that both the scale of solar neutrino oscillations, $m_{\rm sol}
\equiv \sqrt{\Delta m^2_{\rm sol}}\simeq 8\times 10^{-3}\,{\rm eV}$, and the
scale of atmospheric neutrino oscillations, $m_{\rm atm} \equiv
\sqrt{\Delta m^2_{\rm atm}} \simeq 0.05\,{\rm eV}$, are much larger
than the equilibrium neutrino mass $m_*$. Hence, the parameter range
suggested by neutrino oscillations, $m_{\rm sol}\lesssim \mt \lesssim
m_{\rm atm}$, lies entirely in the strong washout regime where
theoretical uncertainties are small and the efficiency factor is still
large enough to allow for successful leptogenesis.

\subsection{CP asymmetry}

The $C\!P$ asymmetry $\varepsilon_1$ is the second crucial
ingredient needed to calculate the baryon asymmetry. To leading
order in the Yukawa coupling $h$, the $C\!P$ asymmetry is
determined by the interference between tree level and vertex plus
self-energy one-loop diagrams \cite{fps95,crv96} and can be consistently 
extracted from the $l\phi \rightarrow l\phi$ scattering processes
\cite{bp98}.

We will be interested in the maximal $C\!P$ asymmetry for given
neutrino masses. Assuming a hierarchy of the heavy neutrinos,
$M_{2,3} \gg M_1$, $\ve_1^{\rm max}$ depends on $M_1$, $\mt$, $m_1$
and $m_3$ \cite{bdp03}. It can be expressed as 
\begin{equation}\label{hmy}
\ve_1^{\rm max}(M_1,\mt,m_1,m_3) = \ve_1^{\rm max}(M_1)\,
\beta(\mt,m_1,m_3)\;, \quad \beta \leq 1 \;.
\end{equation}
The maximal asymmetry, i.e. $\beta=1$, is reached for $m_1=0$ and,
with $m_3=\sqrt{m^2_{\rm atm}+m_1^2}$, it is given by \cite{hmy02}
\begin{equation}\label{e1maxM1}
\ve_1^{\rm max}(M_1)={3\over 16\pi}\,{M_1\,m_{\rm atm}\over v^2}
\simeq 10^{-6}\,\left({M_1\over 10^{10}\,{\rm GeV}}\right) \,
\left({m_{\rm atm}\over 0.05\,{\rm eV}}\right) \; .
\end{equation}
An improved bound is obtained in the limit $m_1/\mt \rightarrow
0$, where one obtains \cite{di02},
\begin{equation}
\beta(m_1) = {m_3 - m_1\over m_{\rm atm}} \;.
\end{equation}
Also important is the case
of quasi-degenerate neutrinos, $m_3 \simeq m_1$. In this region
one finds \cite{hlx03,pdb04},
\begin{equation}\label{beta}
\beta(\mt,m_1) \simeq {m_3-m_1\over m_{\rm atm}}\,
\sqrt{1-{m_1^2\over\mt^2}}  \; .
\end{equation}
For all neutrino mass models with moderately hierarchical heavy
neutrinos, such that Eq.~(\ref{hmy}) applies, the observed baryon
asymmetry yields constraints on the three neutrino parameters
$\mt,m_1$ and $M_1$. In the following two sections we follow
mostly the discussion in Ref.~\cite{bdp04}.

\subsection{Lower bounds on heavy neutrino masses and reheating temperature}

The {\em maximal baryon asymmetry} $\eta_B^{\rm max}(M_1,\mt,m_1)$
is the asymmetry corresponding to $\ve_1^{\rm max}(M_1,\mt,m_1)$.
The CMB bound then amounts to the requirement
\begin{equation}\label{CMBc}
\eta_B^{\rm max}(M_1,\mt,m_1) \geq \eta_B^{CMB}\;.
\end{equation}
This represents an interesting constraint on the space of
the three parameters $M_1$, $m_1$ and $\mt$. We have seen that the absolute
maximum of the $C\!P$ asymmetry is obtained for $m_1=0$. For $m_1\neq 0$
the function $\beta$ suppresses the $C\!P$ asymmetry \cite{di02}.
Furthermore, the $\Delta L=2$ washout term becomes stronger when the absolute
neutrino mass scale increases (cf.~(\ref{DW})). Therefore, the maximal baryon
asymmetry $\eta_B^{\rm max}(M_1,\mt,m_1)$ is maximized for $m_1=0$,
and in this case the allowed region in the space of the parameters $M_1$
and $\mt$ is maximal \cite{bdp02}.

In this way one finds an important lower bound on the value of $M_1$
\cite{di02,bdp02}, by inserting the expression (\ref{e1maxM1})
into the CMB constraint (\ref{CMBc}) (cf.~(\ref{etak})),
\begin{eqnarray}\label{lbM1}
M_1 \geq M_1^{\rm min} &=& {1\over d}\, {16\,\pi\over 3}\,{v^2\over
m_{\rm atm}}\,
{\eta_B^{CMB}\over\k_{\rm f}} \NO\\
&\simeq& 6.4\times 10^{8}\,{\rm GeV} \left(\eta_B^{CMB}\over
6\times 10^{-10}\right) \left(0.05\,{\rm eV}\over m_{\rm
atm}\right)\,\k_{\rm f}^{-1} \;.
\end{eqnarray}
For {\em thermal initial abundance}, and in the limit
$\mt/m_* \rightarrow 0$, one has by definition $\k_f=1$, and therefore
\begin{equation}\label{M1lbza}
M_1 \geq (6.6\pm 0.8)\times 10^8\,{\rm GeV}
\gtrsim 4\times 10^8\,{\rm GeV} \, .
\end{equation}
Here the last inequality is the $3\,\sigma$ bound, with the
experimental value of Eq.~(\ref{etaBCMB}) for $\eta_B^{CMB}$,
and $m_{\rm atm}=(0.051\pm 0.004)\,{\rm eV}$.
In the case of a {\em dynamically generated $N_1$ abundance}
the maximal efficiency
factor is $\k_{\rm f}\simeq 0.18$, which yields the more stringent bound
\begin{equation}\label{M1lbta}
M_1\, \geq\, (3.6\pm 0.4) \times 10^9\,{\rm GeV}\,
\gtrsim \, 2 \times 10^9\,{\rm GeV} \; .
\end{equation}

The most interesting case corresponds to the range
$m_{\rm sol}\lesssim \mt \lesssim m_{\rm atm}$,
for which the power law Eq.~(\ref{plaw}) for $\k_{\rm f}$ can be used.
Using the central value of $\kappa_f$ and neglecting the theoretical
uncertainty, one obtains
\begin{equation}\label{M1lbsw}
M_1 \gtrsim   (3.3\pm 0.4) \times 10^{10}\,{\rm GeV}\,
\left(\widetilde{m}_1\over 10^{-2}\,{\rm eV} \right)^{1.1} \;,
\end{equation}
which yields the $3\,\sigma$ bound
\begin{equation}\label{M1b3s}
M_1 \gtrsim  (1.5 - 10)\times 10^{10}\,{\rm GeV} \; .
\end{equation}
The lower bound on $M_1$  is particularly interesting since it can be
translated into a lower bound on the initial temperature $T_{\rm i}$
which, within inflationary models, corresponds to the reheating
temperature.

So far we have assumed that the temperature $T_{\rm i}$ is larger
than $M_1$. If one relaxes this assumption, the final efficiency factor
is in general reduced. For small $\mt$, however, the threshold value
for $T_{\rm i}$,
below which the reduction is appreciable, is given by $M_1$ itself.
Below this temperature the $N_1$ abundance is either Boltzmann suppressed,
for thermal initial abundance, or the $N_1$ production is considerably
suppressed, for zero initial abundance. Therefore, for values
$\mt\lesssim 10^{-3}\,{\rm eV}$, the bounds (\ref{M1lbza}) and
(\ref{M1lbta}) apply also to the reheating temperature $T_{\rm i}$.

In the more interesting case of strong washout, about $90\%$
of the final baryon asymmetry is produced in a temperature interval
$z_B - 2 \lesssim z \lesssim z_B + 2$. Hence the reheating temperature can
be about $(z_B-2)$ times lower than $M_1$ without any appreciable change in
the predicted final asymmetry \cite{bdp04}. In the interesting range
$m_{\rm sol} \lesssim \mt \lesssim m_{\rm atm}$ one has 
$z_B \simeq 6 - 8$, and therefore the bound Eq.~($\ref{M1b3s}$) gets
relaxed by a factor $4 - 6$, such that
\begin{equation}\label{T1lbsw}
T_{\rm i}\gtrsim (4\times 10^9 - 2\times 10^{10})\,{\rm GeV} \, .
\end{equation}
Compared to the small $\mt$ range, $\mt \lesssim 10^{-3}$~eV, the
lower bound on the reheating temperature is slightly more restrictive
in the favoured range $m_{\rm sol} \lesssim \mt \lesssim m_{\rm atm}$
due to the loss in efficiency.

\subsection{Upper bound on light neutrino masses}

For large values of the absolute neutrino mass scale the $\Delta W$ washout
term cannot be neglected. The final efficiency factor can be calculated
using the approximation that $\Delta W$ starts to be effective for $z>z_B$,
where the asymmetry generation from decays has already terminated. This
works very well in the strong washout regime. Since $\mt\geq m_1$, this does
not introduce any restriction if
$m_1\gtrsim m_* \simeq 10^{-3}\,{\rm eV}$.
One then has
\begin{equation}\label{kfzb}
\k_{\rm f}(\mt,\,M_1\mb^2)\simeq\kappa_{\rm f}(\mt) \,e^{- {\omega\over
z_B}\, \left({M_1\over 10^{10}\,{\rm GeV}}\right)\,
\left({\mb\over {\rm eV}}\right)^2}\, ,
\end{equation}
where $\omega\simeq0.186$ and $\k_{f}(\mt)$ is the efficiency factor
calculated in the regime of small neutrino masses, neglecting the
$\Delta W$ term. Because of the assumption of strong washout one can
use the simple power law Eq.~(\ref{plaw}).  At certain peak values of
$M_1$ and $\mt$ the maximal baryon asymmetry $\eta_B^{\rm max}$
reaches the absolute maximum
\begin{equation}
{\eta_{B}^{\rm peak}(\mb)\over \eta_B^{CMB}}
\propto {\chi\,m_*\,\xi \over \mb^4} \;,
\end{equation}
with $\chi \simeq 1.6\,{\rm eV}^3$ and $\xi\propto m_{\rm atm}^2$.
This finally yields the leptogenesis bound on the neutrino masses,
$m_i \lesssim 0.1\,$eV \cite{bdp03}. Deviations of the quantity $\xi$ 
from unity account for a change of the input parameters as
well as various corrections, such as a possible enhancement of the
$C\!P$ asymmetry or supersymmetry. 

A more precise calculation has to take into account the dependence of
neutrino masses on the renormalization scale. The running of the
atmospheric neutrino mass scale $m_{\rm atm}$ from the Fermi scale
$\mu = m_Z$ to the high scale $\mu \sim T_B$ makes the bound less
restrictive.  On the other hand the bound on $m_i$ obtained at high
energies has to be evolved down to low energies. This second effect is
dominant and thus taking the scale dependence into account makes
the neutrino mass bound more restrictive \cite{akx03}. The smallest
effect is obtained for a Higgs mass $M_h \simeq 150\,{\rm GeV}$, which
leads to a $\sim 20\%$ more restrictive bound \cite{akx03}. Taking
into account this effect one then obtains the $3\sigma$ bound
\cite{bdp04}
\begin{equation}\label{mbound}
m_i \lesssim 0.12\,{\rm eV}\,\xi^{1/4}  \, .
\end{equation}
Note, that the bound $m_i < 0.1\,$eV conservatively accounts for the
theoretical uncertainties including spectator processes \cite{bp01}
which make the bound more restrictive by about $0.02\,$eV \footnote{
In Ref.~\cite{gnx04} the upper bound 0.15~eV has been obtained, which
is 0.03~eV weaker than the bound (\ref{mbound}). About 0.02~eV of this
difference is due to the different treatment of radiative corrections
which depend on the top and Higgs masses.
The remaining 0.01~eV reflects differences in the treatment of thermal
corrections. This is included in the theoretical uncertainty of the
efficiency factor $\k_{\rm f}(\mt)$ (cf.~Eq.~(\ref{plaw})).}.  
The strong suppression of
the baryon asymmetry with increasing neutrino mass scale,
$\eta_B\propto 1/\mb^4$, which is reflected in the dependence
$\xi^{1/4}$,  makes the bound rather stable. This is different from
the lower bounds on $M_1$ and $T_i$ which relax as $1/\xi$
\cite{bdp03}.

It is important to keep in mind that the neutrino mass bound can be
evaded, with some effort. A measurement of the neutrino
mass scale above the leptogenesis bound would require significant
modifications of the minimal leptogenesis scenario that we described.
The possibilities include quasi-degenerate heavy neutrinos
\cite{bdp03,hlx03}, non-thermal leptogenesis scenarios 
\cite{ls91,my94,gpx99}
or a non-minimal seesaw mechanism as in theories with Higgs triplets 
\cite{hs04,rod04,ak04}.

\subsection{Leptogenesis conspiracy}

The upper bound on the neutrino masses (\ref{mbound}) arises
when the information on $m_{\rm atm}$ from neutrino mixing
experiments is employed. The value of $m_{\rm atm}$ sets the scale for the
transition from a hierarchical, with $m_1\ll m_{\rm atm}$, to a
quasi-degenerate neutrino mass spectrum with $m_1\gtrsim m_{\rm
atm}$. The joint action of the $C\!P$ asymmetry suppression for
$m_1\gtrsim m_{\rm atm}$ and $\mt\approx m_1$ (cf. (\ref{beta})),
together with the washout from the $\Delta W$ term, place a limit
to the level of degeneracy of the light neutrino masses and,
using the measured value of the atmospheric neutrino mass
scale, an upper bound on the absolute neutrino mass scale.

We now want to study how this upper bound gets relaxed if the
experimental measurement of the atmospheric neutrino mass scale is
ignored. Since the maximal asymmetry is obtained for hierarchical
neutrinos, i.e. $\b=1$, we have to use the simple bound (\ref{e1maxM1}). 
This also implies that the upper bound on the absolute neutrino mass 
scale will coincide with an upper bound on the atmospheric neutrino mass 
scale itself, since $\mb \simeq m_{\rm atm} \simeq m_3$.
The maximal baryon asymmetry is then approximately given by 
\begin{equation}
\eta_{B}^{\rm max}(M_1,\mt,m_3) = d\,\ve_1^{\rm max}(M_1)\,
\k_{\rm f}(\mt,M_1\,m_3^2) \;. 
\end{equation} 
Using Eq.~(\ref{kfzb}) it is easy to see that the
maximum of the asymmetry is realized for 
\begin{equation}\label{M1} 
M_1 \simeq
2\,z_B\,\,10^{13}\,{\rm GeV}\, \left({0.05\,{\rm eV}\over  m_3}\right)^2\;, 
\end{equation} 
and is then given by 
\begin{equation}\label{etaBmax}
\eta_B^{\rm max}(\mt,m_3)\simeq 0.7\times
10^{-5}\,z_B(\mt)\,\k_{\rm f}(\mt)\, 
\left({0.05\,{\rm eV}\over m_3}\right) \, . 
\end{equation}

In the case of zero initial $N_1$ abundance the maximum is
obtained for $\mt\simeq 2\times 10^{-3}\,{\rm eV}$, where
$k_{\rm f}\simeq 0.2$ and $z_B\simeq 2$ \footnote{This can be inferred
from Fig.~4 and Fig.~5 of Ref.~\cite{bdp04}.}, implying 
\begin{equation}\label{peakm3}
\eta_B^{\rm peak}(m_3)\simeq 3\times 10^{-6}\,
\left({0.05\,{\rm eV}\over m_3}\right) \, . 
\end{equation} 
Note that
the peak lies in the strong washout regime where results do not
depend on the initial conditions. 

For $m_3 \simeq m_{\rm atm}\simeq 0.05\,{\rm eV}$, Eq.~(\ref{peakm3})
is in good agreement with the numerical results of Ref.~\cite{bdp02}. 
If we do not make use of the experimental
information on $m_{\rm atm}$ and just require that the peak asymmetry
is larger than the observed value given by Eq.~(\ref{etaBCMB}), then we
obtain the upper bound 
\begin{equation} 
m_3 \lesssim 250\;{\rm eV} \; .
\end{equation} 
Together with Eq.~(\ref{M1}) this implies the lower bound for  
heavy neutrinos masses,
\begin{equation}
M_1\gtrsim 2\times 10^6\;{\rm GeV} \;.
\end{equation}
This exercise shows that without the experimental
knowledge of $m_{\rm atm}$ the bound on light neutrino 
masses would have been much looser.
However, it also demonstrates, even more remarkably, that the neutrino 
oscillation data, together with the laboratory bounds on light neutrino
masses, represents a highly non trivial test of thermal leptogenesis.

As discussed in the previous section, the leptogenesis bound of 0.1~eV 
appears to have a theoretical uncertainty of about 0.03~eV. For comparison,
ten years ago it was believed, based on the same equations, that Majorana
masses $m_3 \sim 10$~keV and $M_1 \sim 1$~TeV were compatible with thermal
leptogenesis \cite{Luty}. During the past years theory and experiment
contributed about equally, on a logarithmic scale, to the progress. 
On the theoretical side the better understanding of the Boltzmann equations
was important whereas the measurement of the atmospheric neutrino mass scale
was the crucial experimental ingredient. 

Actually, the successful matching of thermal leptogenesis predictions
with experimental data is even more intriguing. From Eq.~(\ref{etaBmax}) 
we can derive an upper bound on $\mt$ by using the strong washout behaviour 
$\k_{\rm f}\simeq 2/(z_B\,K)$ (cf.~(\ref{kfth})) and imposing the CMB bound,
which yields, with $m_{\rm atm} = \sqrt{m_3^2-m_1^2}$,
\begin{equation}
\mt \lesssim 20\,{\rm eV}\,
\left({0.05\,{\rm eV}\over m_{\rm atm}}\right) \; , 
\end{equation} 
again in very good agreement with the numerical results \cite{bdp02}. 
From this expression one reads off that only for 
$m_{\rm atm}\lesssim 1\,{\rm eV}$ it is possible to have 
$\mt\sim m_{\rm atm}$. If the experiments had found
$m_{\rm atm}\gg 1\,{\rm eV}$, thermal leptogenesis would have worked only for 
models with $\mt\ll m_{\rm atm}$, requiring a large amount of fine tuning, 
as already pointed out. Furthermore, the favoured strong washout regime
implies $m_{\rm atm} > m_*$. Hence, leptogenesis favours for the atmospheric
mass scale the range $10^{-3}\,{\rm eV} - 1\,{\rm eV}$, in remarkable
agreement with experimental data. 
Thanks to this `conspiracy' the seesaw mechanism,
for the same values of the involved parameters, explains equally
well both the neutrino masses and the observed baryon asymmetry, with
a remarkable independence on the assumptions about the
inflationary stage or, more generally, the cosmological stages
that precede leptogenesis.

This conspiracy would become even more impressive if the lightest
neutrino mass should turn out to be larger than $m_*$. This would
imply, in a completely model independent way, $\mt\gtrsim m_*$, with
thermal leptogenesis in the strong washout regime. Together with the
upper bound $m_1<0.1\,{\rm eV}$, this selects the optimal 
{\em leptogenesis window} 
$10^{-3}\,{\rm eV}\lesssim m_1 \lesssim 0.1\,{\rm eV}$
for the absolute neutrino mass scale.

\section{Flavour aspects}

Leptogenesis is closely related to other processes involving neutrinos
and charged leptons. The upper bound on the light neutrino masses and
the lower bound on the heavy neutrino masses have already been
discussed in the previous section. Very interesting are also the
connection with neutrino mixing and with lepton flavour changing
processes in supersymmetric theories. During the past years these
subjects have been studied in great detail by many groups
(cf.~\cite{mas02,xin03,kin03,af04,ell04}).  In the following we shall
discuss two important examples, the possible connection between $C\!P$
violation in neutrino oscillations and leptogenesis, i.e. at low and
high energies, and the relation between leptogenesis and the heavy
neutrino mass spectrum.

In the standard model with right-handed neutrinos the masses and mixings
of leptons are described by three complex matrices,
\begin{equation}
{\cal L}_M = \overline{\nu_L} m_D \nu_{R} +
        \overline{e}_{L} m_l e_{R}\; +
        \overline{\nu_R^c} M \nu_{R}\; + \; h.c.\;.
\end{equation}
For the Majorana mass matrix $M$ one expects $M \gg m_D$, which leads to 
three light and three heavy Majorana mass eigenstates,
$\nu \simeq \nu_L + \nu_L^c = \nu^c$ and  $N \simeq \nu_R + \nu_R^c = N^c$.

In the following we will work in a basis where $M$ is diagonal and real,
with $M_1 < M_2 < M_3$, which is appropriate for leptogenesis. The light
neutrino mass matrix $m_\nu$ and the charged lepton mass matrix $m_l$
are then diagonalized by the unitary transformations,
\begin{eqnarray}
V^{(\nu)\dagger} m_\nu V^{(\nu)*} &=&
- \left(\begin{array}{ccc} m_1 & 0 & 0 \\
  0 & m_2 & 0 \\ 0 & 0 & m_3 \end{array}\right)\,
=\,m_\nu^{\rm diag} \;, \\
V^{(e)\dagger} m_e \widetilde{V}^{(e)} &=&
\left(\begin{array}{ccc} m_e & 0 & 0 \\
  0 & m_\mu & 0 \\ 0 & 0 & m_\tau \end{array}\right)\,=\,m_l^{\rm diag}\;.
\end{eqnarray}
The leptonic (MNS) mixing matrix
\begin{equation}
U = V^{(e)\dagger} V^{(\nu)}\;
\end{equation}
describes the couplings of mass eigenstates in the charged current,
\begin{equation}
{\cal L}^{(l)}_{EW} = -{g\over\sqrt{2}}
\overline{e}_{L} \gamma^\mu U \nu_{L}\ W^-_\mu \ + \ldots \;,
\end{equation}
which leads to neutrino oscillations.

It is well known that, in general, $C\!P$ violation in neutrino oscillations
and in leptogenesis are unrelated \cite{bp96}. The reason is that the $C\!P$
asymmetries
$\varepsilon_i$ in heavy neutrino decays only depend on $m_D^\dagger m_D$.
Hence, changing $m_D$ to $K m_D$, where $K$ is a general unitary matrix,
leaves $\varepsilon_i$ invariant whereas the leptonic mixing matrix
$U = V^{(e)\dagger} V^{(\nu)}$
is changed to $U = V^{(e)\dagger} K V^{(\nu)}$, and therefore arbitrary.
Still, the question remains whether in
some physically well motivated cases a connection between $C\!P$ violation
at low and high energies exists.

$C\!P$ violating observables are most conveniently described by weak basis
invariants which are inert under a unitary transformation,
$l\rightarrow K l$, of the lepton doublet $l= (\nu_L,e_L)$.
For neutrino oscillations the appropriate variable is the
commutator between the hermitian matrices $H_\nu = m_\nu m_\nu^\dagger$
and $H_l = m_l m_l^\dagger$ \cite{bmx01,bra03},
\begin{eqnarray}
{\rm Tr}[H_\nu,H_l]^3\, \propto\,
\Delta_{\nu 21} \Delta_{\nu 32} \Delta_{\nu 31}
\Delta_{l 21} \Delta_{l 32} \Delta_{l 31}\,J_l \;,\\
\qquad \Delta_{\nu 21} = m_2^2-m_1^2\,, \ldots , 
\Delta_{l 21} = m_\mu^2-m_e^2\,, \ldots \;,
\end{eqnarray}
where $J_l$ is the leptonic Jarlskog invariant \cite{jar85},
\begin{equation}
J_l\, =\, \rm{Im}[U_{11}U_{22}U_{12}^*U_{21}^*]\, \propto\, \sin{\delta}\;,
\end{equation}
which is proportional to the $C\!P$ violating phase $\delta$ of the
mixing matrix $U$.
In a basis where the charged lepton matrix $m_l$ is diagonal and real
one has
\begin{equation}\label{cpnu}
{\rm Tr}[H_\nu,H_l]^3\, \propto\,
\Delta_{l 21} \Delta_{l 32} \Delta_{l 31}\,
{\rm Im}[H_{\nu12}H_{\nu23}H_{\nu31}]\;;
\end{equation}
correspondingly, for diagonal and real $m_\nu$ one has
\begin{equation}\label{cpl}
{\rm Tr}[H_\nu,H_l]^3\, \propto\,
\Delta_{\nu 21} \Delta_{\nu 32} \Delta_{\nu 31}\,
{\rm Im}[H_{l12}H_{l23}H_{l31}]\;.
\end{equation}

The $C\!P$ asymmetry in $N_1$ decays can be conveniently expressed in
terms of the weak basis invariant \cite{bf00}
\begin{equation}\label{lepcp}
\varepsilon_1 \propto {\rm Im}[m_D^\dagger m_\nu m_D^*]_{11}\;.
\end{equation}
Comparing Eqs.~(\ref{cpl}) and (\ref{lepcp}) the independence of $C\!P$
violation at low and high energies is obvious. In a basis where $m_\nu$ is
diagonal and real, $C\!P$ violation in neutrino oscillations is
entirely determined by the phases $m_l$ whereas the lepton asymmetry
only depends on the phases of $m_D$.

An interesting example is the case of only two heavy Majorana neutrinos,
$N_1$ and $N_2$, which corresponds to the limit where $N_3$ decouples
\cite{fgy02}. For a given texture of $m_D$ one then obtains 
(cf.~(\ref{cpnu}), (\ref{lepcp})),
\begin{equation}
\ve_1 \propto
{\rm Im}[H_{\nu12}H_{\nu23}H_{\nu31}] \propto \sin {\delta}\;.
\end{equation}
Hence both, the $C\!P$ violation at low and at high energies, are
determined by the Dirac phase $\delta$. A further low energy quantity
is the Majorana phase entering neutrinoless double beta decay. In 
some models with maximal atmospheric neutrino mixing this phase
coincides with the leptogenesis phase \cite{gl03}. All $C\!P$ phases 
can also be related in models of spontaneous $C\!P$ violation 
\cite{bmx01,ach04}. 

Another important question is the connection between leptogenesis and the
heavy neutrino mass spectrum. In the lepton mass eigenstate basis
the neutrino mass matrix can be reconstructed from data, i.e. the leptonic
mixing matrix and the neutrino masses,
\begin{equation}
m_\nu = - U^{(\nu)} m_\nu^{\rm diag} U^{(\nu)T} \;.
\end{equation}
The heavy neutrino mass matrix is then determined by $m_\nu$ and $m_D$,
\begin{equation}
M^{-1} = - m_D^{-1} m_\nu (m_D^T)^{-1}\;.
\end{equation}
Using the SO(10) mass relation $m_D = m_u$, where $m_u$ denotes the up-type
quark mass matrix, and assuming that in the lepton mass eigenstate basis
$m_u$ is also diagonal and real, $m_u = {\rm diag}(m_u,m_c,m_t)$, one
obtains the heavy neutrino masses in terms of $m_\nu$, $m_u$, $m_c$ and
$m_t$ \cite{afs03}. Neglecting the hierarchy among the small neutrino
masses one estimates that the hierarchy among the heavy neutrinos is
very large,
\begin{equation}
{M_1 \over M_3} \sim \left({m_u \over m_t}\right)^2 \sim 10^{-10} \;.
\end{equation}
With $M_3 \sim m_t^2/\sqrt{\Delta m^2_{\rm atm}} \sim 10^{15}$~GeV,
this implies $M_1 \sim 10^5$~GeV. A detailed study yields masses in the
range $M_1 \sim 10^4 \ldots 10^6$~GeV. As we saw in the previous section,
such small masses are incompatible with conventional thermal leptogenesis.
A possible way out are quasi-degenerate heavy neutrino masses \cite{afs03}.
Alternatively, the most naive SO(10) mass relations are not correct.

Parameters consistent with thermal leptogenesis have recently been obtained
in a six-dimensional SO(10) GUT model, compactified on an orbifold
\cite{abc03}. Due to mixings with a heavy lepton doublet and a heavy
right-handed down-type quark one finds for the mass matrices in a particular
flavour basis the relations,
\begin{equation}
M \propto m_u\;, \quad m'_D \sim m'_d \sim m'_e\;.
\end{equation}
Here $M$ and $m_u$ are diagonal $3\times 3$ matrices, 
and $m'_D$, $m'_d$ and $m'_e$
are $4\times 4$ matrices, due to the mixing with the heavy states.
Integrating them out, the neutrino mass matrix can be expressed
approximately in terms of the quark mass matrices,
\begin{equation}
m_\nu \propto  m_d {1\over m_u} m_d^T\;.
\end{equation}
For the light neutrino masses this leads to the estimate
\begin{equation}
{m_1\over m_3} \sim \left({m_d\over m_b}\right)^2 {m_t\over m_b} \sim 0.1\;.
\end{equation}
A detailed calculation \cite{abc03} confirms this result. Furthermore,
one finds the mixing angle $\Theta_{13} \sim (m_c m_b)/(m_t m_s) \sim 0.1$,
and $\varepsilon \sim 10^{-6}$, $\widetilde{m}_1 \sim 10^{-2}$~eV,
$M_1 \sim (m_u/m_t) M_3 \sim 10^{10}$~GeV, consistent with thermal
leptogenesis.

\section{Cosmological aspects}

It is well known that the temperature required by thermal leptogenesis,
$T_B \gtrsim 10^9$~GeV, is potentially in conflict with the thermal
production of gravitinos in the early universe \cite{kl84,ekn84}.
Late gravitino decays
after nucleosynthesis significantly alter the successful BBN predictions.

The production of gravitinos is dominated by QCD processes, and the gravitino
number density increases linearly with the reheating temperature after
inflation,
\begin{equation}
{n_{3/2}\over n_\gamma} \propto {g_3^2\over M_{\rm p}^2}\,T_R\;,
\end{equation}
where $g_3$ is the QCD gauge coupling.
Correspondingly, the BBN upper bound on the allowed gravitino energy density,
$\rho_{3/2}=m_{3/2}n_{3/2}$, implies an upper bound on the reheating
temperature. Detailed studies \cite{kkm01,cex03} lead to the stringent bounds
$T_R < 10^7 - 10^9$~GeV for gravitino masses in the range
$m_{3/2} = 0.1 - 1$~TeV. Hence, unstable gravitinos are in conflict with
thermal leptogenesis, unless the gravitino mass is very large,
$m_{3/2} \gtrsim 50$~TeV.

Non-thermal leptogenesis models are still compatible with the above bounds on
the reheating temperature. For instance, in some supersymmetric models the
scalar partner $\widetilde{N}_1$ of the heavy neutrino $N_1$ might be the
inflaton \cite{msx93} and its decays could then generate the baryon asymmetry.
In this case the leptogenesis temperature can be below $10^7$~GeV
\cite{ery04,gnx04}, which would be consistent with the above gravitino
bounds on $T_R$. However, a recent analysis of the BBN constraints \cite{kkm04}
\footnote{Note that the analysis strongly depends on the assumed 
primordial $^6$Li abundance.}
with particular attention to the hadronic decay modes of the gravitino
yields the much stronger bound $T_R \ll 10^6$~GeV for gravitino
masses $m_{3/2} = 0.1 - 1$~TeV. Hence, unless the gravitino is extremely heavy,
also non-thermal leptogenesis appears to be inconsistent with unstable
gravitinos.

Already in connection with thermal leptogenesis, it has therefore been
suggested that the gravitino may be the lightest superparticle (LSP) and
stable \cite{bbp98}. In this case, gravitino production is
enhanced \cite{mmy93},
\begin{equation}
{n_{3/2}\over n_\gamma} \propto {g_3^2\over M_{\rm p}^2}\,
\left({m_{\widetilde{g}}\over m_{3/2}}\right)^2\,T_R\;,
\end{equation}
where $m_{\widetilde{g}}$ is the gluino mass. Consistency with BBN and the
observed amount of dark matter then yields an upper bound on the gravitino
mass and a lower bound on the mass of the next-to-lightest superparticle
(NSP). In \cite{bbp98} the case of a higgsino NSP was analyzed, which is
now disfavoured due to the improved BBN bounds on hadronic NSP decays
\cite{kkm04}. Still viable is the case where a scalar lepton is the
NSP \cite{fiy04}. For $T_B = 3\times 10^9$~GeV the upper bound on the
gluino mass is $m_{\widetilde{g}} < 1.3$~TeV if a scalar tau is the NSP; for
a scalar neutrino NSP one finds $m_{\widetilde{g}} < 1.8$~TeV \cite{fiy04}.

The relic density of gravitinos is determined by thermal production at the
reheating temperature $T_R$ after inflation and also by NSP decays after
their freeze-out temperature. It is an interesting possibility that the
latter process dominates. This would be the case for low reheating
temperatures, incompatible with leptogenesis. One then obtains
a prediction for the amount of gravitino dark matter which is independent
of the reheating temperature \cite{frt03}. If gravitinos are the only
component of dark matter, the superparticles have to be rather heavy.
For $\widetilde{\tau}_R$ as NSP one finds $m_{3/2}=0.2-1$~TeV and
$m_{\widetilde{\tau}_R} \geq 0.5$~TeV \cite{fst04}.

\begin{figure}[t]
\centerline{\psfig{figure=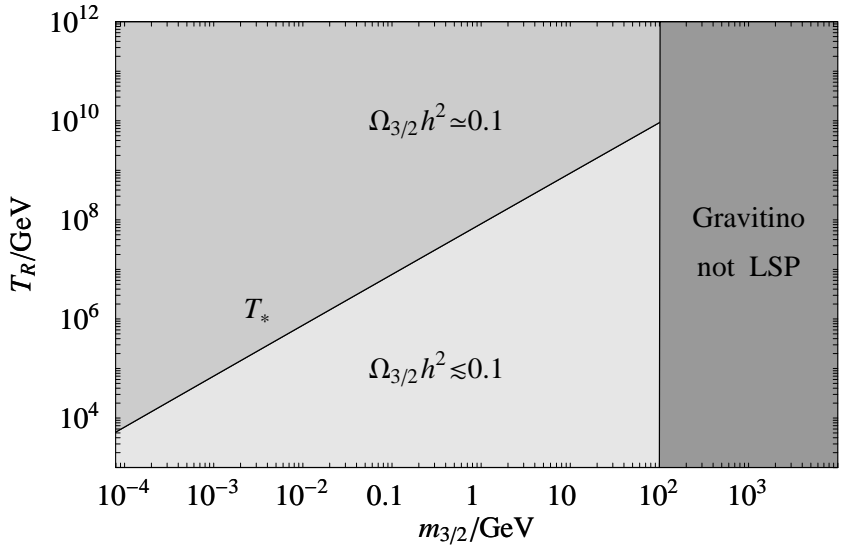,width=11cm}}
\caption{Relic gravitino density for different values of reheating
temperature and gravitino mass. $\xi/\eta^2 =1$.
$m_{\widetilde{g}}=1$~TeV, which implies
$m_{3/2} < 0.1$~TeV for a stable gravitino. For $T_R > T_*$,
$\Omega_{3/2}h^2$ is independent of $T_R$ and $m_{3/2}$. }
\end{figure}

Recently, it has been pointed out that the thermal production of gravitinos
is significantly changed in theories where the gauge coupling depends on
the expectation value of a scalar field $\phi$ \cite{bhr03}. For instance,
in the case of gaugino mediation one has,
\begin{equation}\label{gtherm}
{1\over g_0^2} + {\phi_T\over M} = {1\over g^2(\phi_T)}\;,
\end{equation}
where $g_0$ is the zero-temperature gauge coupling, $M$ is a mass scale
between the unification scale and the Planck mass, and $\phi_T$ is the
deviation of the field $\phi$ from its zero-temperature value at
temperature $T$. At temperatures above a critical temperature,
\begin{equation}
T_* \sim m_{3/2} \left({M_{\rm p} \over m_{\widetilde{g}}}\right)^{1/2} \;,
\end{equation}
the gauge couplings decreases, and the gravitino production is frozen. 
Remarkably,
the relic gravitino density is essentially determined just by the gluino mass
\cite{bhr03},
\begin{eqnarray}\label{final}
  \Omega_{3/2}h^2 \simeq (0.05 - 0.2) \times
  \left(\frac{m_{\widetilde{g}}}{1~{\rm TeV}}\right)^{3/2}
  \left(\frac{\xi}{\eta^2}\right)^{1/4}\;,
\end{eqnarray}
where the factor $\xi/\eta^2$ 
\footnote{Here the parameter $\xi^{1/2}$ denotes the ratio of dilaton mass 
and gravitino mass.}
 depends on the mechanism of supersymmetry
breaking.
For gaugino  and gravity mediation one has $\xi/\eta^2 = {\cal O}(1)$.
The observed amount of dark matter is then obtained for a gluino mass
$m_{\widetilde{g}} \sim 1$~TeV, which will be tested at the LHC. As
Fig.~5 illustrates, the temperature where thermal leptogenesis takes place
is likely to be larger than the critical temperature $T_*$. For a gluino
mass $m_{\widetilde{g}} \sim 1$~TeV one then obtains automatically the
observed amount of cold dark matter.

Finally, thermal leptogenesis can also be consistent with gravitino dark
matter if the gravitino is very light, $m_{3/2} \simeq 0.1-10$~MeV
\cite{fy02,fiy042}, which is realized in gauge mediation models.
Alternatively, the gravitino can also be very heavy,
$m_{3/2} \sim 100$~TeV \cite{ls03,ikx04}, as in anomaly mediation.

In summary, leptogenesis strongly constrains the nature of dark matter.
For very heavy unstable gravitinos, $m_{3/2} \sim 100$~TeV,
where ordinary WIMPs can be the dark matter, the superparticle mass spectrum
is strongly restricted. Alternatively, the gravitino has to be the LSP with
a mass below $\sim 0.1$~TeV; the observed value of $\Omega_{CDM}h^2$
can then be naturally explained. Hence, the identification of the invisible
dark matter, which will hopefully take place at colliders in the coming
years, will also shed light on baryogenesis and therefore on the origin of 
the visible matter.

\section*{Acknowledgement}

We would like to thank K.~Hamaguchi and M.~Ratz for helpful
discussions and comments.

\end{document}